\documentclass[twoside,twocolumn,english,aps,psb,superscriptaddress]{revtex4}
\usepackage[T1]{fontenc}
\usepackage[latin1]{inputenc}
\usepackage{amsmath}
\usepackage{graphicx}

\makeatletter

\providecommand{\LyX}{L\kern-.1667em\lower.25em\hbox{Y}\kern-.125emX\@}

\usepackage{babel}
\makeatother
\begin{document}

\title{Effects of Density of States Asymmetry on Boundary Resistance of
Ferromagnetic-Nonferromagnetic Metal Interface}

\author{Pavel Krotkov}

\affiliation{National High Magnetic Field Laboratory, Florida State University,
Tallahassee, Florida 32310, USA}

\affiliation{L.D. Landau Institute for Theoretical Physics, Russian Academy of
Sciences, 2 Kosygina st., 117334 Moscow Russia}

\author{Maxim Dzero%
\footnote{Present address: U.S. Department of Energy Ames Laboratory, Ames,
IA 50011. E-mail: dzero@ameslab.gov.%
}}

\affiliation{National High Magnetic Field Laboratory, Florida State University,
Tallahassee, Florida 32310, USA}

\begin{abstract}
Asymmetry of conduction of electrons in the spin majority and minority
bands of a ferromagnetic metal is well known to produce spin accumulation
at the boundary with a normal metal when a current is injected through
the interface. However, little emphasis has been put on the density
of states (DOS) asymmetry in diffusive ferromagnetic-nonferromagnetic
multilayers. We found that if DOS and conduction asymmetry differ,
the electric potential in a ferromagnet falls off to its bulk value
exponentially on a scale of the spin diffusion length. Therefore the
boundary contribution to resistance dramatically depends on whether
the potential difference is measured close to the interface or farther
than the spin diffusion length from it. This result is not altered
by taking surface resistance or spin-flips on the intermetallic boundary
into account. Explicit answers for common multilayered spin-valve
structures are given.

PACS number(s): 72.25.Pn, 72.25.Ba, 73.40.Cg, 73.40.Jn
\end{abstract}
\maketitle
\newcommand{\tsf}{\tau_{\uparrow\downarrow}}

\newcommand{\lsd}{l_{\mathrm{sd}}}

\newcommand{\mfp}{\lambda}

\newcommand{\si}{\rho}

\newcommand{\ca}{\alpha}

\newcommand{\ados}{\beta}

\newcommand{\f}[1]{\mbox {\boldmath\(#1\)}}

\newcommand{\ts}[1]{\textstyle #1}

\newcommand{\dif}[1]{[#1]}

\newcommand{\ave}[1]{\{#1\}}

\newcommand{\const}{\mathop{\mathrm{const}}\nolimits}

\section{Introduction}

The possibility to exploit electron spin to manipulate currents in
electronic microdevices has generated wide interest in the rapidly
expanding field of spin electronics \cite{Awschalom02,Prinz98,Gregg02}.
Technically, the itinerant electron spin may be put to use if the
spin-flip scattering, intermixing the spin-up and spin-down channels,
occurs at a rate $\tau _{\uparrow \downarrow }^{-1}$ which is slow
compared to the inverse mean free time $\tau ^{-1}$ which characterizes
the dissipative processes in charge of electrical resistivity. Separation
of electrons in two channels is a key element of the physics of a
central spintronics phenomenon of giant magneto-resistance (GMR) of
metallic ferromagnetic-nonferromagnetic metal multilayers. 

The two-channel description of diffusive electrical transport from
a ferromagnetic (FM) to a nonferromagnetic metal was given in \cite{Sohn}
and on a thermodynamic basis also in \cite{Johnson}. Due to the exchange
splitting of the Fermi surface in a FM metal the Fermi energy corresponds
to different parts of the band structure for spin-up and spin-down
electrons, and therefore the two types of electrons have different
densities of state $N_{\uparrow ,\downarrow }$ and conductions $\sigma _{\uparrow ,\downarrow }$.
Because of the conduction asymmetry \begin{equation}
\ca =\frac{\sigma _{\uparrow }-\sigma _{\downarrow }}{\sigma _{\uparrow }+\sigma _{\downarrow }}\label{eq:ca}\end{equation}
an electrical current passing from a FM to a nonferromagnetic metal
is polarized and injects magnetization. This effect shows up as an
additional contact resistance.

The size $\lsd $ of the region of the spin accumulation is set by
the equilibrium between spin injection through the interface and the
spin-flipping processes in the bulk of the paramagnet. The spin diffusion
length $\lsd $ relates to the mean free path $\mfp $ as $\lsd /\mfp \sim (\tsf /\tau )^{1/2}$.
An analysis of the two-channel model was given by Valet and Fert \cite{Valet}
on the basis of the Boltzmann transport equation in the limit $\lsd \gg \mfp $.
They have also applied the model to the geometry of the GMR with current
perpendicular to the plane (CPP-GMR). A significant advantage of their
approach is the ability to describe measurable quantities in detail
in terms of parameters of the theory, which allows to extract the
latter from experiment \cite{Dubois,Pratt,Piraux}.

The formulas in \cite{Valet}, however, have been derived in a flawed
assumption that while the conductivities $\sigma _{\uparrow ,\downarrow }$
of the majority and minority spins are different, the corresponding
densities of states (DOS), $N_{\uparrow ,\downarrow }$, are equal.
But the difference in $N_{\uparrow ,\downarrow }$ is essential as
it leads to a jump of the electric potential on an intermetallic interface,
which also was not observed in the later work \cite{fert96}. Neglect
of the difference in $N_{\uparrow ,\downarrow }$ results in an incorrect
variation of the electric potential with distance from an intermetallic
interface, and changes, \emph{inter alia}, the boundary contribution
to resistance.

The aim of the present paper is to provide expressions for boundary
resistances of common multilayered systems corrected to allow the
measurements of potential drop to be performed at any distance from
interfaces, not necessarily $\gg \lsd $. Correlative results for
an array of domain walls have been obtained in \cite{Gor'kov}. In
Sec. \ref{sec:Basic-Equations} a review of the two-channel diffusive
transport theory is given, whereas in Sec. \ref{sec:Results} we demonstrate
our results for a single ferromagnetic-nonferromagnetic interface
and for typical spin-valve structures. Sec. \ref{sec:Conclusions}
summarizes the findings and provides a brief discussion. Finally,
an appendix with details of our approach to calculation is presented
at the end of the paper.

\section{Basic Equations\label{sec:Basic-Equations}}

Starting from the Boltzmann equation, Valet and Fert \cite{Valet}
showed that in the limit $\lsd \gg \mfp $ one can describe transport
in the system introducing electrochemical potentials $\mu _{\si }$
for the two spin channels $\si =\uparrow ,\downarrow $ as sums of
chemical potentials (accounting for the kinetic energy of electrons)
and the potential energy. For small deviations from equilibrium the
chemical potential is given by the non-equilibrium electron density
$n_{\si }$ divided by the density of states $N_{\si }$ on the Fermi
surface:

\begin{equation}
\mu _{\si }=n_{\si }/N_{\si }-eU.\label{eq:muupdown}\end{equation}
Here $-e$ is the electron charge, $U$ is the electric potential,
and the condition $n_{\uparrow }=-n_{\downarrow }$ is fulfilled owing
to electroneutrality. 

We now briefly review the equations of two-channel diffusive transport
to establish notations. Electron current $\mathbf{j}_{\si }$ in the
channel $\si $ is a sum of a diffusion component and a contribution
from the electron drift under the influence of an electric field $\mathbf{E}=-\f \partial U$:\begin{equation}
\mathbf{j}_{\si }=-(-e)D_{\si }\f \partial n_{\si }+\sigma _{\si }\mathbf{E}=\sigma _{\si }\f \partial \mu _{\si }/e,\label{eq:j}\end{equation}
where we used the Einstein relation $\sigma _{\si }=e^{2}N_{\si }D_{\si }$
between the conductivity $\sigma _{\si }$ and the diffusion constant
$D_{\si }$. In the bulk the excess electron density $n_{\uparrow }$
is zero, the electrochemical potentials for the two channels are equal:
$\mu _{\uparrow }=\mu _{\downarrow }$, and the current polarization
is $\ca $ (\ref{eq:ca}). Relaxation of non-equilibrium polarization
of current near a boundary between two materials due to spin-flip
processes is described by the continuity equations \begin{equation}
\f \partial \mathbf{j}_{\si }=en_{\si }/\tsf .\label{eq:dj}\end{equation}

It is convenient to introduce the divergence $\mu _{s}\equiv \mu _{\uparrow }-\mu _{\downarrow }$
between the two electrochemical potentials. It is non-zero only in
the presence of spin accumulation: from (\ref{eq:muupdown}) one finds
that it is related to $n_{\uparrow }$ via \begin{equation}
\mu _{s}=n_{\uparrow }(N_{\uparrow }^{-1}+N_{\downarrow }^{-1}).\label{eq:musdef}\end{equation}
 Substituting (\ref{eq:musdef}) into (\ref{eq:j}) and (\ref{eq:dj})
we find that $\mu _{s}$ satisfies the diffusion equation 

\begin{equation}
\f \partial ^{2}\mu _{s}=\mu _{s}/\lsd ^{2},\label{eq:mus}\end{equation}
where the spin diffusion length\begin{equation}
\lsd =\sqrt{\frac{\tsf }{e^{2}}\frac{\sigma _{\uparrow }^{-1}+\sigma _{\downarrow }^{-1}}{N_{\uparrow }^{-1}+N_{\downarrow }^{-1}}}.\end{equation}
Equation (\ref{eq:mus}) by means of relation (\ref{eq:musdef}) describes
the exponential decay of non-equilibrium spin density away from a
boundary. 

In order to find currents $\mathbf{j}_{\si }$ in each channel and
the electric potential $U$ we introduce 

\begin{equation}
\mu _{0}=\frac{\sigma _{\uparrow }\mu _{\uparrow }+\sigma _{\downarrow }\mu _{\downarrow }}{\sigma },\label{eq:mu0}\end{equation}
where $\sigma =\sigma _{\uparrow }+\sigma _{\downarrow }$ is the
net conductivity. Differentiating $\mu _{0}$ yields $\mathbf{j}=\sigma \f \partial \mu _{0}/e$,
where $\mathbf{j}=\mathbf{j}_{\uparrow }+\mathbf{j}_{\downarrow }$
is the net current being conserved throughout the sample ($\f \partial \mathbf{j}=0$).
The spatial distribution of $\mu _{0}$ is found from $\f \partial ^{2}\mu _{0}=0$,
hence in one dimension \begin{equation}
\mu _{0}(x)=\const +ejx/\sigma .\label{eq:mu0sol}\end{equation}
The spin current $\mathbf{j}_{s}=\mathbf{j}_{\uparrow }-\mathbf{j}_{\downarrow }$
may then be found from the known distribution of $\mu _{s}$:\begin{equation}
\mathbf{j}_{s}=\alpha \mathbf{j}+\sigma (1-\ca ^{2})\f \partial \mu _{s}/2e.\label{eq:js}\end{equation}
From the definition (\ref{eq:muupdown}) we find that the spatial
distribution of $U$ is given by

\begin{equation}
-eU=\frac{N_{\uparrow }\mu _{\uparrow }+N_{\downarrow }\mu _{\downarrow }}{N_{\uparrow }+N_{\downarrow }}=\mu _{0}+\frac{\ados -\ca }{2}\mu _{s},\label{eq:-eU}\end{equation}
where we have introduced the asymmetry of the density of states\begin{equation}
\ados =\frac{N_{\uparrow }-N_{\downarrow }}{N_{\uparrow }+N_{\downarrow }}.\end{equation}
Equations (\ref{eq:musdef}), (\ref{eq:mus}), (\ref{eq:js}), and
(\ref{eq:-eU}) completely describe electron transport inside a layer. 

At distances $\gg \lsd $ from boundaries, when the divergence $\mu _{s}$
decays to zero, the two electrochemical potentials $\mu _{\uparrow ,\downarrow }=\mu _{0}+\frac{1}{2}(\pm 1-\alpha )\mu _{s}$
as well as the electric potential (\ref{eq:-eU}) collapse to $\mu _{0}$.
I.e. $\mu _{0}$ is the value of the electrochemical potential that
would be realized were there not non-equilibrium distribution of spin.
However, if measurements are done closer than $\lsd $ to an interface,
one has to take into account that the non-equilibrium divergence of
$\mu _{\uparrow ,\downarrow }$ has not yet subsided. 

To find charge and current distribution through the whole multilayer
sample, Eqs. (\ref{eq:musdef}), (\ref{eq:mus}), (\ref{eq:js}),
and (\ref{eq:-eU}) have to be supplemented by boundary conditions.
In the absence of an interface resistance and spin-flip scattering,
the appropriate boundary conditions are the continuity of $\mu _{\uparrow ,\downarrow }$
and $\mathbf{j}_{\uparrow ,\downarrow }$ \cite{Valet}. The continuity
of $\mu _{\uparrow ,\downarrow }$ shows that $\mu _{0}$ and $U$
are discontinuous at intermetallic interfaces with the jumps\begin{equation}
\dif \mu _{0}=\ts \frac{1}{2}\dif {\ca }\mu _{s},\qquad -e\dif U=\ts \frac{1}{2}\dif {\ados }\mu _{s},\label{eq:difmu0}\end{equation}
where $\dif {\ca }$ is the jump in the conduction asymmetry (\ref{eq:ca})
and $\dif {\ados }$ is the jump in the asymmetry of the density of
states. The discontinuity in $\mu _{0}$ appears in transport measurements
farther than $\lsd $ from an interface as a boundary contribution
to resistances\begin{equation}
R_{I\infty }=\dif \mu _{0}/ej=\dif {\ca }\mu _{s}/2ej.\label{eq:Rinfty}\end{equation}

Discontinuities of the electrochemical potentials and the spin current
may be introduced \cite{fert96} to account for interface resistance
and spin-flip scattering. One then has to allow for the jumps of $\mu _{\uparrow ,\downarrow }$
in (\ref{eq:difmu0}), which affects the answer (\ref{eq:Rinfty}).
But the inclusion of new parameters for interface resistance and spin-flip
scattering renders the formulas very cumbersome, not changing the
principal point of the paper. Therefore, we have delegated the discussion
of the interface resistance and spin-flip scattering to Appendix.

\section{Results\label{sec:Results}}

Equation (\ref{eq:-eU}) presents the spatial variation of the electric
potential, provided $\mu _{0}$ and $\mu _{s}$ for the problem in
question are found. As revealed by its second term, if $\beta -\alpha \ne 0$,
spin accumulation near an intermetallic interface causes $U$ to collapse
to its bulk asymptotic value $-\mu _{0}/e$ on a spin diffusion length
scale. Hence, the form of the boundary contribution to resistance
has to be modified dependent on how close to the boundary the measurements
take place. 

\begin{figure}
\includegraphics[  width=0.95\columnwidth]{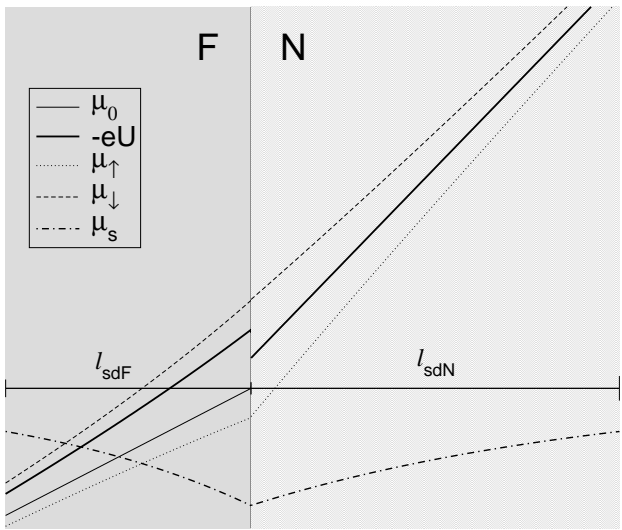}

\caption{Sketch of spatial variation of potentials at the boundary between
ferromagnetic and normal metals.\label{cap:Sketch-of-spatial}}
\end{figure}

Anyhow, since the jump $-e\dif U$ at an interface as given by the
second of Eqs. (\ref{eq:difmu0}) differs, generally speaking, from
the jump of $\dif \mu _{0}$, if measured in the immediate vicinity
to the interface, instead of (\ref{eq:Rinfty}) the interface resistance
would rather be \begin{equation}
R_{I0}=-\dif U/j=\dif {\ados }\mu _{s}/2ej.\label{eq:R0}\end{equation}
Note that the jumps $-e\dif U$ and $\dif \mu _{0}$ not only differ
in absolute value, but may in principle have opposite signs. 

At intermediate distances from an interface the boundary resistance
must be obtained from (\ref{eq:-eU}): $R_{x_{1}x_{2}}=\left(U(x_{1})-U(x_{2})\right)/j$
by a complete solution of the problem for the whole multilayer structure.

\subsection{A Single Ferromagnetic-Nonferromagnetic Metal Interface}

The solution is straightforward in the simplest case of an isolated
boundary at the origin between two different materials with constants
denoted by the two subscripts 1 and 2. We here have decided to illustrate
our claim with more general than simply a ferromagnetic-nonferromagnetic
interface example at an expense of only slightly more complex expressions.

Equating spin currents (\ref{eq:js}) we find $\mu _{s}$ at the origin
and reestablish the Valet-Fert expression for the boundary resistance:\begin{equation}
R_{I\infty }=\frac{(\ca _{1}-\ca _{2})^{2}}{\lambda _{1}+\lambda _{2}}.\label{eq:isolated}\end{equation}
In this expression and below we use the shorthand notation\begin{equation}
\lambda =\sigma (1-\ca ^{2})/\lsd .\end{equation}
For arbitrary $x_{1}<0$, $x_{2}>0$ the resistance between the two
points is \begin{equation}
R_{x_{1}x_{2}}=|x_{1}|/\sigma _{1}+x_{2}/\sigma _{2}+R_{I12},\label{eq:17}\end{equation}
where the first two terms are the series resistances of the two materials
over the lengths $x_{1}$ and $x_{2}$ respectively, and the interface
contribution is \begin{equation}
R_{I12}=R_{I\infty }\left(1-\frac{(\ca _{1}-\ados _{1})e^{x_{1}/\lsd {}_{1}}-(\ca _{2}-\ados _{2})e^{-x_{2}/\lsd {}_{2}}}{\ca _{1}-\ca _{2}}\right).\label{eq:18}\end{equation}
This expression exponentially decays from the value \begin{equation}
R_{I0}=R_{I\infty }\frac{\ados _{1}-\ados _{2}}{\ca _{1}-\ca _{2}}\label{eq:20}\end{equation}
 very close to the boundary to $R_{I\infty }$ farther than $\lsd $
for it. 

Note that if one of the materials is a nonferromagnetic metal, then
the corresponding exponential term in (\ref{eq:18}), which describes
decay of the electric potential in it, is zero. This is because in
a normal metal the electric potential (\ref{eq:-eU}) falls onto the
asymptote $\mu _{0}$ right after the interface, not exponentially
with the distance from it.

Position dependence of the electric potential $-eU$ and of the electrochemical
potentials $\mu _{0}$, $\mu _{\uparrow }$, $\mu _{\downarrow }$,
$\mu _{s}$ for an isolated boundary between a ferromagnetic ($\alpha _{\mathrm{F}}=0.5$,
$\beta _{\mathrm{F}}=-0.5$) and a normal metal ($\alpha _{\mathrm{N}}=\beta _{\mathrm{N}}=0$)
is plotted in Fig. \ref{cap:Sketch-of-spatial}. Note that because
$\beta _{\mathrm{F}}$ is taken negative, the jumps $-e\dif U$ and
$\dif \mu _{0}$ have opposite signs. 

Allowing for interface resistance and spin-flips complicates the formulas
(\ref{eq:isolated}), (\ref{eq:18}), and (\ref{eq:20}) without changing
the qualitative conclusion of the difference between $R_{I0}$ and
$R_{I\infty }$. Expressions for boundary resistances in this case
are provided in Appendix.

\subsection{Spin-valve Structures}

\begin{figure}
\includegraphics[  width=0.95\columnwidth]{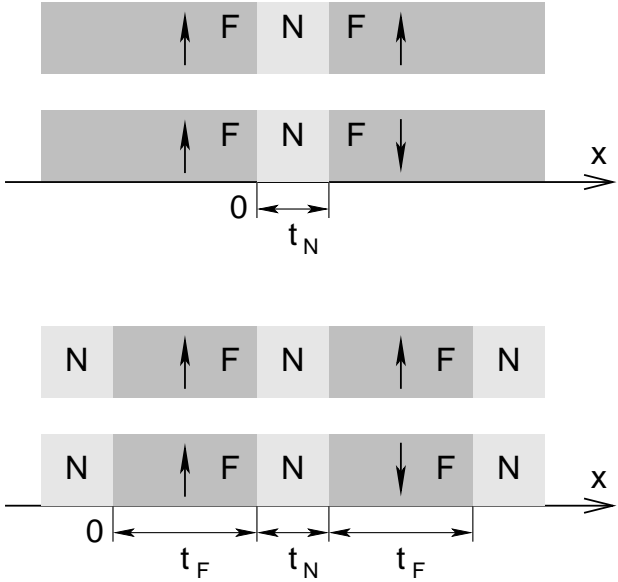}

\caption{Sketch of a spin-valve multilayer structure, with supplying wires
(two bottom structures) and without them. In each case parallel and
antiparallel configurations of the two ferromagnetic layers are depicted.
\label{cap:multi}}
\end{figure}

In the end, we cite formulas for a simplest spin-electronic device
--- a trilayer `spin-valve': a (thin) layer of paramagnetic metal
of thickness $t_{\mathrm{N}}$ sandwiched between two ferromagnetic
layers with parallel or antiparallel orientations of the magnetic
momenta. We consider both the cases with and without supplying normal
metal electrodes (see Fig. \ref{cap:multi}). In the first case we
assume the two ferromagnetic layers to have equal thickness $t_{\mathrm{F}}$.

In case of a symmetric trilayer without the supplying electrodes (the
uppermost structure in Fig. \ref{eq:isolated}) the Valet-Fert boundary
resistances of the two interfaces are \begin{equation}
R_{I\infty }=\frac{\ca ^{2}}{\lambda _{\mathrm{F}}+\lambda _{\mathrm{N}}\coth \theta _{\mathrm{N}}/2},\label{eq:R3s}\end{equation}
where $\theta _{\mathrm{N}}=t_{\mathrm{N}}/\lsd {}_{\mathrm{N}}$.
In the limit of large $t_{\mathrm{N}}\gg \lsd {}_{\mathrm{N}}$ this
expression turns into (\ref{eq:isolated}) for an isolated boundary.
When the thickness of the paramagnetic layer $t_{\mathrm{N}}\to 0$,
we obtain that $R_{I\infty }$ vanishes as $R_{I\infty }\approx \ca ^{2}t_{\mathrm{N}}/2\sigma _{\mathrm{N}}$. 

In what follows we omit the series contributions to resistance like
the first two term in (\ref{eq:17}). For any structure in Fig. \ref{cap:multi}
these contributions are trivial. We thus simply cite the boundary
part of the total resistance of a symmetric trilayer between two probe
points $x_{L}<0$ and $x_{R}>t_{\mathrm{N}}$:\begin{equation}
R_{IRL}=R_{I\infty }\left(2-\frac{\ca -\ados }{\ca }\left(e^{x_{L}/\lsd {}_{\mathrm{F}}}+e^{(t_{\mathrm{N}}-x_{R})/\lsd {}_{\mathrm{F}}}\right)\right).\label{eq:22}\end{equation}
For the measurement points very close to the boundaries $x_{L}=0$
and $x_{R}=t_{\mathrm{N}}$ (\ref{eq:22}) reduces to \begin{equation}
R_{I0}=2\ados R_{I\infty }/\ca .\end{equation}

For an antisymmetric trilayer without the supplying electrodes (the
second structure from the top in Fig. \ref{cap:multi}), when the
right ferromagnetic layer has the opposite conduction asymmetry $-\ca $,
the Valet-Fert boundary resistances are \begin{equation}
R_{I\infty }=\frac{\ca ^{2}}{\lambda _{\mathrm{F}}+\lambda _{\mathrm{N}}\textrm{th }\theta _{\mathrm{N}}/2}.\label{eq:R3a}\end{equation}
When $t_{\mathrm{N}}\to 0$ the resistance tends to a finite value
$R_{I\infty }=\ca ^{2}\lsd {}_{\mathrm{F}}/(1-\ca ^{2})\sigma _{\mathrm{F}}$,
thus justifying the term `spin-valve'. The boundary contribution to
resistance between two arbitrary points $x_{L}<0$ and $x_{R}>t_{2}$
is\begin{equation}
R_{IRL}=R_{I\infty }\left(2-\frac{\ca +\ados }{\ca }e^{(t_{\mathrm{N}}-x_{R})/\lsd {}_{\mathrm{F}}}+\frac{\ca -\ados }{\ca }e^{x_{L}/\lsd {}_{\mathrm{F}}}\right)\end{equation}
and tends to \begin{equation}
R_{I0}=2(\ca -\ados )R_{I\infty }/\ca \end{equation}
 when $x_{L}=0$ and $x_{R}=t_{\mathrm{N}}$.

\begin{widetext}

For a symmetric five-layer with two supplying normal-metal electrodes
(the third structure from the top in Fig. \ref{cap:multi}) the boundary
resistances of the two external and the two internal interfaces are
respectively\begin{eqnarray}
R_{Ie\infty } & = & \ca ^{2}\frac{\lambda _{\mathrm{N}}\coth \theta _{\mathrm{N}}/2+\lambda _{\mathrm{F}}\textrm{th }\theta _{\mathrm{F}}/2}{\lambda _{\mathrm{F}}^{2}+\lambda _{\mathrm{N}}\lambda _{\mathrm{F}}\coth \theta _{\mathrm{F}}(1+\coth \theta _{\mathrm{N}}/2)+\lambda _{\mathrm{N}}^{2}\coth \theta _{\mathrm{N}}/2},\label{eq:R5se}\\
R_{Ii\infty } & = & \ca ^{2}\frac{\lambda _{\mathrm{N}}+\lambda _{\mathrm{F}}\textrm{th }\theta _{\mathrm{F}}/2}{\lambda _{\mathrm{F}}^{2}+\lambda _{\mathrm{N}}\lambda _{\mathrm{F}}\coth \theta _{\mathrm{F}}(1+\coth \theta _{\mathrm{N}}/2)+\lambda _{\mathrm{N}}^{2}\coth \theta _{\mathrm{N}}/2}.\label{eq:R5si}
\end{eqnarray}

For an antisymmetric five-layer (the lowermost structure in Fig. \ref{cap:multi})

\begin{eqnarray}
R_{Ie\infty } & = & \ca ^{2}\frac{\lambda _{\mathrm{N}}\textrm{th}\theta _{\mathrm{N}}/2+\lambda _{\mathrm{F}}\textrm{th }\theta _{\mathrm{F}}/2}{\lambda _{\mathrm{F}}^{2}+\lambda _{\mathrm{N}}\lambda _{\mathrm{F}}\coth \theta _{\mathrm{F}}(1+\textrm{th}\theta _{\mathrm{N}}/2)+\lambda _{\mathrm{N}}^{2}\textrm{th}\theta _{\mathrm{N}}/2},\label{eq:R5ae}\\
R_{Ii\infty } & = & \ca ^{2}\frac{\lambda _{\mathrm{N}}+\lambda _{\mathrm{F}}\textrm{th }\theta _{\mathrm{F}}/2}{\lambda _{\mathrm{F}}^{2}+\lambda _{\mathrm{N}}\lambda _{\mathrm{F}}\coth \theta _{\mathrm{F}}(1+\textrm{th}\theta _{\mathrm{N}}/2)+\lambda _{\mathrm{N}}^{2}\textrm{th}\theta _{\mathrm{N}}/2}.\label{eq:R5ai}
\end{eqnarray}
The boundary contribution to resistance between two arbitrary points
$x_{L}<0$ and $x_{R}>t_{\mathrm{N}}+2t_{\mathrm{F}}$ is for both
symmetric and antisymmetric five-layered structures simply the total
boundary resistance of all the interfaces $R_{IRL}=2(R_{Ii\infty }+R_{Ie\infty })$
since in the normal-metal supplying electrodes the electric potential
coincides with its bulk asymptote $-\mu _{0}/e$ everywhere after
the interface.

\end{widetext}

\section{Conclusions\label{sec:Conclusions}}

In this brief study we focused on the effects of the density of states
asymmetry in the diffusive transport through multilayered ferromagnetic-nonferromagnetic
metal structures. 

Whereas conduction asymmetry brings about spin accumulation in a layer
of thickness of the order of the spin diffusion length, and determines
the overall drop of the electric potential in this layer (apart from
the trivial series resistance contributions), the DOS asymmetry results
in a jump of the electric potential directly at an intermetallic interface.
In a standard Valet-Fert theory only the boundary contributions to
resistance $R_{I\infty }$ from the overall potential drop in the
spin accumulation layer are calculated. These boundary resistances
emerge only when measurements are done farther than the spin diffusion
length from interfaces.

However, the series resistances that build up over the distance $\sim \lsd $
between the measurement points turn out to be larger than the boundary
resistance $R_{I\infty }$. To identify the part of the resistance
from the current conversion process on the boundary it is advantageous
to attach probes closer to the interface. But in that case it should
be remembered that because of the asymmetry of the density of states
the interface contribution to the resistance differs from $R_{I\infty }$.

\section{Acknowledgments}

The authors are indebted to Prof. Lev P. Gor'kov for innumerable enlightening
discussions and guidance through the work, and acknowledge support
from DARPA through the Naval Research Laboratory Grant No. N00173-00-1-6005. 

\appendix

\section*{General Solution of the Equations\label{sec:General-Solution-of}}

In this appendix we expound the general approach we used to calculate
the spatial distribution of potentials in a multilayered structure.
The diffusion equation (\ref{eq:mus}) in an interval $a<x<b$ in
one dimension has a solution\begin{equation}
\mu _{s}(x)=\frac{\mu _{s}(a)\sinh \frac{x-a}{\lsd }+\mu _{s}(b)\sinh \frac{b-x}{\lsd }}{\sinh \frac{b-a}{\lsd }}\label{eq:musx}\end{equation}
that falls off exponentially on a length scale $\lsd $ from the values
$\mu _{s}(a)$ and $\mu _{s}(b)$ at the boundaries $a$ and $b$. 

First consider the boundaries with no interface resistance or spin-flip
scattering. In this case the boundary conditions of continuity of
the electrochemical potentials and of the currents apply. The form
(\ref{eq:musx}) for a solution in each layer with respective $\lsd $
relies on the values of $\mu _{s}$ on the boundaries, which thus
need to be put equal on the two sides of each interface to satisfy
the boundary condition of the continuity of $\mu _{s}=\mu _{\uparrow }-\mu _{\downarrow }$.
The second boundary condition for the electrochemical potentials leads
to the expression (\ref{eq:difmu0}) for the jump of $\mu _{0}$,
which together with solution (\ref{eq:mu0sol}) in each layer with
respective $\sigma $ yields answer for the spatial distribution (\ref{eq:mu0sol})
of $\mu _{0}$. 

Substituting (\ref{eq:musx}) into (\ref{eq:js}), we obtain spin
currents at the boundaries:\begin{eqnarray}
j_{s}(a) & = & \ca j+(-B\mu _{s}(a)+A\mu _{s}(b))/2e,\nonumber \\
j_{s}(b) & = & \ca j+(-A\mu _{s}(a)+B\mu _{s}(b))/2e,\label{eq:jsajsb}
\end{eqnarray}
where \begin{equation}
A=\frac{\lambda }{\sinh \frac{b-a}{\lsd }},\qquad B=\lambda \coth \frac{b-a}{\lsd }.\end{equation}
In case of a semi-infinite interval, when either $b\to \infty $ or
$a\to -\infty $, $\mu _{s}$ on the corresponding end tends to zero
and $A=0$, $B=\lambda $. 

Equating (\ref{eq:jsajsb}) on both sides of each interface, we obtain
an algebraic system of equations on the interface values of $\mu _{s}$.
Once found, these values then give boundary resistances (\ref{eq:Rinfty})
and (\ref{eq:R0}). In this way we derived Eqs. (\ref{eq:R3s}), (\ref{eq:R3a}),
(\ref{eq:R5se}), (\ref{eq:R5si}), and (\ref{eq:R5ae}), (\ref{eq:R5ai}).

Interface resistance as well as spin-flip scattering at an interface
may be taken into account by appropriate boundary conditions \cite{fert96}.
Non-zero surface resistance leads to discontinuity of electrochemical
potential proportional to the current in the respective channel and
spin-flip scattering induces discontinuity of the spin current proportional
to the average value of $\mu _{s}$:\begin{equation}
\dif \mu _{\si }=er_{\si }\ave j_{\si },\qquad \dif j_{s}=\frac{1}{er_{\mathrm{sf}}}\ave \mu _{s},\label{eq:hardbc}\end{equation}
where $\ave {\ldots }$ denotes an average of a discontinuous quantity
between the values on the two sides of a boundary. The surface resistances
$r_{\si }$ in each channel may be parametrized by the total resistance
$4r$ and the surface resistance asymmetry \begin{equation}
\gamma =\frac{r_{\uparrow }-r_{\downarrow }}{r_{\uparrow }+r_{\downarrow }}.\end{equation}

Solution (\ref{eq:musx}) of Eq. (\ref{eq:mus}) inside a layer then
obviously still holds, if $\mu _{s}(a)$ and $\mu _{s}(b)$ are understood
as values of $\mu _{s}$ on the internal side of each boundary. Substituting
this into expression (\ref{eq:js}) yields the values of the spin
current on the internal side of each boundary, which may be combined
with the boundary conditions (\ref{eq:hardbc}) to give an algebraic
equation on the interface values of $\mu _{s}$. However, it is easier
to rewrite Eqs. (\ref{eq:musx}) and (\ref{eq:jsajsb}) in terms of
the average values $\ave \mu _{s}$ and $\ave j_{s}$ to produce:\begin{eqnarray}
\ave j_{s}(a) & = & \ca _{r}j+(-B_{r}\ave \mu _{s}(a)+A_{r}\ave \mu _{s}(b))/2e,\nonumber \\
\ave j_{s}(b) & = & \ca _{r}j+(-A_{r}\ave \mu _{s}(a)+B_{r}\ave \mu _{s}(b))/2e\label{eq:36}
\end{eqnarray}
in analogy with (\ref{eq:jsajsb}). Here parameters of interface resistance
$r$, $\gamma $ and of interface spin-flip scattering $r_{\mathrm{sf}}$
now enter only through the renormalized constants $\ca _{r}$, $A_{r}$,
and $B_{r}$:\begin{eqnarray}
\ca _{r} & = & \frac{\ca -(A+B)\gamma r/2}{1+(A+B)r/2},\label{eq:37}\\
A_{r} & = & \frac{A\textrm{(1-}r/2r_{\mathrm{sf}}\textrm{)}}{(1+Br/2)^{2}-(Ar/2)^{2}},\\
B_{r} & = & \frac{B+(B^{2}-A^{2})r/2+(1+Br/2)/r_{\mathrm{sf}}}{(1+Br/2)^{2}-(Ar/2)^{2}}.\label{eq:39}
\end{eqnarray}

The formal analogy between (\ref{eq:36}) and (\ref{eq:jsajsb}) allows
to reuse the system of algebraic equations on the interface values
of $\mu _{s}$ with the substitution of $\ave \mu _{s}$ as variables
and (\ref{eq:37})-(\ref{eq:39}) as parameters. However, the expressions
for the jumps of $\mu _{0}$ and $U$ at the boundaries are more complex
than (\ref{eq:difmu0}):\begin{eqnarray}
\dif \mu _{0} & = & er\left((1+\ave \alpha \gamma )j+(\ave \alpha +\gamma )\ave j_{s}\right)+\ts \frac{1}{2}\dif \alpha \ave \mu _{s},\\
-e\dif U & = & er\left((1+\ave \beta \gamma )j+(\ave \beta +\gamma )\ave j_{s}\right)+\ts \frac{1}{2}\dif \beta \ave \mu _{s}.
\end{eqnarray}
 Hence the boundary resistances (\ref{eq:Rinfty}), (\ref{eq:R0})
will also be different. 

As an example and to give the reader a feel of the complexity of the
formulas that arise when surface resistance and spin-flips are taken
into account we cite our answers for a problem considered earlier
in the text with simple boundary conditions of continuity, i.e. a
single boundary at the origin between two materials. 

\begin{widetext}

The boundary resistances farther than $\lsd $ from the interface
and close to it are respectively

\begin{eqnarray}
R_{I\infty } & = & r(1-\gamma ^{2})+\frac{\dif {\ca }^{2}+r\left(\mfp _{1}(\ca _{2}+\gamma )^{2}+\mfp _{2}(\ca _{1}+\gamma )^{2}\right)+2r_{\mathrm{sf}}^{-1}(\gamma +\ave {\ca })^{2}}{\mfp _{1}+\mfp _{2}+\mfp _{1}\mfp _{2}r+r_{\mathrm{sf}}^{-1}(2+(\mfp _{1}+\mfp _{2})r/2)},\\
R_{I0} & = & r(1-\gamma ^{2})+\frac{\dif {\ca }\dif {\ados }+r\left(\mfp _{1}(\ca _{2}+\gamma )(\ados _{2}+\gamma )+\mfp _{2}(\ca _{1}+\gamma )(\ados _{1}+\gamma )\right)+2r_{\mathrm{sf}}^{-1}(\gamma +\ave {\ca })(\gamma +\ave {\ados })}{\mfp _{1}+\mfp _{2}+\mfp _{1}\mfp _{2}r+r_{\mathrm{sf}}^{-1}(2+(\mfp _{1}+\mfp _{2})r/2)}.
\end{eqnarray}
The dependence of the boundary resistance on the distance from the
interface is given by 

\begin{equation}
R_{I12}=R_{I\infty }+\frac{\ados _{2}-\ca _{2}}{2ej}\mu _{s}(+0)e^{-x_{2}/\lsd {}_{2}}-\frac{\ados _{1}-\ca _{1}}{2ej}\mu _{s}(-0)e^{-x_{1}/\lsd {}_{1}},\end{equation}
where \begin{eqnarray}
\mu _{s}(+0) & = & 2ej\frac{\dif \ca +r\left(\mfp _{1}(\ca _{2}+\gamma )+(\ave {\ca }+\gamma )/4r_{\mathrm{sf}}\right)}{\mfp _{1}+\mfp _{2}+\mfp _{1}\mfp _{2}r+r_{\mathrm{sf}}^{-1}(2+(\mfp _{1}+\mfp _{2})r/2)},\\
\mu _{s}(-0) & = & 2ej\frac{\dif \ca -r\left(\mfp _{2}(\ca _{1}+\gamma )+(\ave {\ca }+\gamma )/4r_{\mathrm{sf}}\right)}{\mfp _{1}+\mfp _{2}+\mfp _{1}\mfp _{2}r+r_{\mathrm{sf}}^{-1}(2+(\mfp _{1}+\mfp _{2})r/2)}
\end{eqnarray}
are the values of $\mu _{s}$ on the two sides of the boundary.

\end{widetext}

\end{document}